\documentclass{ws-ijmpe}
\usepackage[super,compress]{cite}
\usepackage{graphicx,latexsym,amssymb, epsfig}
\usepackage{multirow,amsmath,array,booktabs}
\usepackage{subfigure}
\usepackage{tabularx}
\usepackage{color}
\usepackage{bm}
\usepackage[figuresright]{rotating}
\usepackage[section]{placeins}
\usepackage{slashed}
\usepackage{graphicx}
\usepackage{tikz}

\newcolumntype{R}{>{\raggedleft\arraybackslash}X}

\newcommand{\nc}{\newcommand}       
\nc{\vc}[1] {\mbox{\boldmath $#1$}} 
\nc{\del}       {\partial}              
\nc{\bra}       {\langle}               
\nc{\ket}       {\rangle}               
\nc{\bras}[1]   {\langle #1|}           
\nc{\kets}[1]   {|#1\rangle}            
\nc{\mapleft}[1]{           
	\smash{\mathop{\,          %
			\hbox to 1.5cm{\rightarrowfill}\, }\limits_{#1}}}
\nc{\beq}     {\begin{eqnarray}} \nc{\eeq}    {\end{eqnarray}}
\nc{\nn}      {\\\nonumber} \nc{\vs}      {\vspace{-0.275cm}}
\nc{\fra}    {\frac{1}{2}}
\nc{\mb}        {\mathbf}

\begin{document}

\markboth{Y. Zhang, P. Liu, and J. Hu}{The properties of neutron star from realistic nucleon-nucleon interaction within relativistic Hartree-Fock model}

\catchline{}{}{}{}{}

\title{The properties of neutron star in the framework of relativistic Hartree-Fock model with unitary correlation operator method}

\author{Ying Zhang}

\address{Department of Physics, School of Science, Tianjin University, Tianjin 300072, China\\yzhangjcnp@tju.edu.cn}

\author{Peng Liu}

\address{Tianjin Institute of Aerospace Mechanical and Electrical equipment, Tianjin 300301, China}

\author{Jinniu Hu}

\address{School of Physics, Nankai University, Tianjin 300071,  China\\
	hujinniu@nankai.edu.cn}

\maketitle

\begin{history}
\received{Day Month Year}
\revised{Day Month Year}
\end{history}

\begin{abstract}
The properties of neutron star are studied in the framework of relativistic Hartree-Fock (RHF) model with realistic nucleon-nucleon ($NN$) interactions, i.e. Bonn potentials. The strong repulsion of $NN$ interaction at short range is properly removed by the unitary correlation operator method (UCOM). Meanwhile, the tensor correlation is neglected due to the very rich neutron environment in neutron star, where the total isospin of two nucleons can be approximately regarded as $T=1$. The equations of state of neutron star matter are calculated in $\beta$ equilibrium and charge neutrality conditions. The properties of neutron star, such as mass, radius, and tidal deformability are obtained by solving the Tolman–Oppenheimer–Volkoff equation and tidal equation. The maximum masses of neutron from Bonn A, B, C potentials are around $2.2M_\odot$. The radius are $12.40-12.91$ km at $1.4M_\odot$, respectively. The corresponding tidal deformabilities are $\Lambda_{1.4}=293-355$. All of these properties are satisfied the recent observables from the astronomical and gravitational wave devices and are consistent with the results from the relativistic Brueckner-Hartree-Fock model.

\end{abstract}


\ccode{PACS numbers: 21.65.+f, 21.65.Cd, 24.10.Cn, 21.60.-n}


\section{Introduction}
The gravitational wave signal from two neutron star merger was firstly detected by LIGO and Virgo collaborations in 2017, as GW170817 event, which started the new era of multi-messenger astronomy and impacted many other aspects, such as, astrophysics, nuclear physics, and so on~\cite{abbott17a,abbott17b,abbott17c,goldstein17,coulter17,troja17,haggard17,abbott18a}. The GW170817 event emerged more useful information of neutron star, like its tidal deformability, besides its mass and radius~\cite{hinderer08,hinderer10,fattoyev13,fattoyev18,annala18}. These data provide a good opportunity to understand the structure of neutron star more entirely and constrain the equation of state (EOS) of neutron star matter strongly~\cite{most18,lim18,tews18,zhang18}.  

The EOS of neutron star matter is usually obtained in the framework of nuclear many-body methods with nucleon-nucleon ($NN$) potentials in the conditions of $\beta$ equilibrium and charge neutrality~\cite{oertel17,shen02}. The present available EOSs can be summarized as two types in general. The first type is generated by the density functional theories based on the effective $NN$ interactions, such as, Skyrme Hartree-Fock model~\cite{vautherin72,bender03,stone07}, Gogny Hartree-Fock model~\cite{goriely09}, relativistic mean-field (RMF) model~\cite{serot86,ring96,meng06,liang15}, relativistic Hartree-Fock (RHF) model~\cite{long06,long07,sun08,long12}, and so on. These effective potentials are generated by fitting the qualities of nuclear many-body systems around the saturation density, $\rho_0$, i. e., the ground-state properties of several doubly magic nuclei and the empirical saturation properties of infinite nuclear matter. These density functional theories already can be extended to study the most nuclei in nuclear landscape and their excitation properties~\cite{erler12,xia18}. Therefore, a natural question is whether the EOSs from density functional theories can be applied to investigate the neutron star, whose central density is estimated as larger than $5\rho_0$. It is found that some of them can generate the massive neutron star, while others give the confused behaviors at high density region~\cite{dutra12,dutra14}.   

To reduce the uncertainties of density functional theories, the second type EOSs are obtained by {\it ab initio} nuclear many-body method with realistic $NN$ potentials~\cite{machleidt87,machleidt89,wiringa95,machleidt01,epelbaum15a,epelbaum15b,entem15}, such as variational method,  quantum Monte Carlo method, coupled cluster method, many-body perturbation theory, Brueckner-Hartree-Fock model, and so on~\cite{akmal98,carlson15,dickhoff04,hagen14a,carbone13,carbone14,hagen14,drischler14,drews15,drews16,li06,baldo07}. These methods properly take into account the short range correlation and tensor force in realistic $NN$ potentials, who can arise the similar behaviors of EOSs for the compact matter. However, these many-body methods in nonrelativistic framework cannot reproduce the empirical saturation properties completely. The three-body force or the relativistic effect should be introduced to improve the description on EOSs, such as relativistic Brueckner-Hartree-Fock (RBHF) model~\cite{brockmann90,krastev06,sammarruca10,dalen10,shen19}.

With recent precise measurements on three compact stars, i.e. PSR J1614-2230, PSR J0348+0432, and PSRJ0740+6620, the maximum mass of neutron star should be larger than $2M_\odot$~\cite{demorest10,fonseca16,antoniadis13,cromartie19}.  In GW170817 event, the radius of neutron stars with $1.4M_\odot$ are constrained as less than $13.8$ km and the corresponding tidal deformability is less than $800$~\cite{annala18,most18,lim18,baiotti19}. Actually, it is found that the radii and tidal deformability has a strong correlations in various theoretical models~\cite{fattoyev18}. These latest data brought many more strict constraints of present EOSs. In recent work, Tong {\it et al.} employed the correlations between the tidal deformability and neutron star radius, neutron drop radii and symmetry energy at $2\rho_0$ to extract their detailed values~\cite{tong19}. It is expected that more and more observables of neutron stars will be obtained with the rapid development of technologies.

As we known, the tensor force in neutron star matter should be very weak due to its isospin singlet character. Therefore, it is enough to only treat the short range correlation of realistic $NN$ potentials in neutron-rich matter~\cite{wang12,hu13}. The Jastrow function was included in the RHF model to deal with short range correlations of Bonn A, B, C potentials by us, as the relativistic central variational method for neutron star~\cite{hu11,hu17}. The maximum masses and corresponding radius generated by such method can reproduce the results from RBHF model very well. Furthermore, another very powerful short range correlation scheme, named as unitary correlation operator method (UCOM) was also applied in RHF framework to consider the short range correlation of central force, which also can generate the similar EOS of the neutron-rich matter with that from RBHF model~\cite{hu10}.

In present work, the properties of neutron star, such as maximum mass, radii, and tidal deformability will be investigated in the framework of RHF model with Bonn potentials. To remove their strong repulsion at short range distance, the UCOM will be taken into account. This paper is arranged as follows. In section II, the fundamental formulas of RHF model with UCOM will be given. The numerical results and discussion will be shown in section III. In section IV, the conclusion will be summarized in section IV.

\section{The relativistic Hartree-Fock model with unitary correlation operator method}

Twenty years ago, a very attractive method was developed by Feldmeier {\it et al.} in terms of the unitary correlation operator method (UCOM) to treat the strong repulsion at short range distance and the tensor correlation at intermediate range of nucleon-nucleon ($NN$) interaction~\cite{feldmeier98,neff03}. The UCOM was demonstrated extremely efficient to provide the reasonable binding energies and wave functions of light nuclei with the modified Afnan-Tang force ~\cite{feldmeier98}.  The essence of UCOM is to introduce a unitary transformation,
\beq
\psi=U\phi.
\eeq
Here, $\psi$ indicates the exact wave function of many-body system, while $\phi$ presents the uncorrelated trivial wave function. Hence, if the details of  correlation function, $U$ are well known, we can obtain the exact wave function in terms of the trivial wave one. The unitary correlation operator $U$ is written as
\beq
U=\exp\{-iC\},~~~~C=C^\dag,
\eeq
where $C$ is an hermitian generator of the correlations. Hence, we can get a set of new equation of motion for nucleon as
\beq
&&\mathcal H\psi=E\psi,\nn
&&U^\dag\mathcal H U\psi=E\psi.
\eeq

Furthermore, $C$ includes a two-body operator or many-body operator because a one-body operator would only cause the unitary transformations of single particle states,
\beq
C=\sum^A_{i<j} c(i,j)+\text{three-body}+...
\eeq
However, we make an approximation of UCOM up to two-body correlation terms \cite{feldmeier98}. This approximation is justified for the short-range correlation, because the probability that three-body nucleons enter their interaction ranges is small at the normal nuclear matter density. Now, we can use the two-body correlation operator $u(i,j)$ instead of generalized correlation operator $U$. For a Hamiltonian in nuclear matter, which consists of a one-body kinetic energy and a two-body potential,
\beq
\mathcal H=\sum^A_i T_i+\sum^A_{i<j}V(i,j),
\eeq
after the modification of UCOM, it will be changed as
\beq
\widetilde{\mathcal H}&=&u^\dag(i,j)\mathcal H u(i,j)\nn
&=&\sum^A_i T_i+\sum^A_{i<j}\widetilde V(i,j),
\eeq
where an effective two-body interaction $\widetilde V(i,j)$ is generated from the short range correlation on two-body kinetic energy and bare $NN$ interaction,
\beq
\widetilde V(i,j)=u^\dag(i,j)V u(i,j)+u^\dag(i,j)(T_i+T_j) u(i,j)-(T_i+T_j).
\eeq
This effective interaction including the short range correlation plays a same role as the $G$-matrix in Brueckner method~\cite{li06,baldo07}.

In actual calculation, it is not convenient to directly adopt the operator form. This correlator, $u(i,j)$, can be expressed in terms of a coordinate transformation $R_+(r)$ for the radial distance,
\beq\label{ucomfun}
R_+(r)=r+\alpha\left( \frac{r}{\beta} \right)^{\eta} \exp(-\exp(r/\beta)).
\eeq
The parameter $\alpha$ determines the overall amount of the shift and $\beta$ the length scale. $\eta$ controls the steepness around $r=0$. The double-exponential can ensure the correlator just has effect in the short-range distance. The most common operators in Hamiltonian are transferred with following in the framework of UCOM,
\beq
u^\dag(i,j) r u(i,j)&=& R_+(r)\nn
u^\dag(i,j) V(r) u(i,j)&=& V(R_+(r))\nn
u^\dag(i,j) p_r u(i,j)&=& \frac{1}{\sqrt{R'_+(r)}}\frac{1}{r}p_r \frac{1}{\sqrt{R'_+(r)}},
\eeq
where $p_r$ is the radial momentum, $\bra \vec r|p_r|\phi\ket=-i\frac{\partial}{\partial r}\bra \vec r|\phi\ket$. The other operators related with angular momentum, as $\vec l$ and $\vec s$ are unchanged.

The Lagrangian of Bonn potential, which will be used in this work can be written as \cite{machleidt89},
\beq\label{bonnlag}
\mathcal{L}_{\text{int}} &=& \bar{\psi} \bigg[- g_{\sigma}\sigma-g_\delta\tau_a\delta^a-\frac{f_\eta}{m_\eta}\gamma_{5}\gamma_{\mu}\partial^{\mu}{\eta}- \frac{f_\pi}{m_\pi}\gamma_{5}\gamma_{\mu}{\tau_{a}}
\partial^{\mu}{\pi^{a}}\nn
&-& g_{\omega}\gamma_{\mu}\omega^{\mu}+\frac{f_\omega}{2M}\sigma_{\mu\nu}
\partial^\nu\omega^{\mu }
-g_\rho\gamma_\mu\tau_a\rho^{a\mu }+\frac{f_\rho}{2M}\sigma_{\mu\nu}
\partial^\nu\tau_a\rho^{a\mu }\bigg]{\psi} \nn
& + &\frac{1}{2}\partial_{\mu}\sigma\partial^{\mu}\sigma
- \frac{1}{2}m_{\sigma}^2\sigma^2
+\frac{1}{2}\partial_{\mu}\delta^a\partial^{\mu}\delta^a
- \frac{1}{2}m_{\delta}^2\delta^{a2}\nn
&+& \frac{1}{2}\partial_{\mu}\eta\partial^{\mu}\eta
- \frac{1}{2}m_{\eta}^2{\eta}^2
+ \frac{1}{2}\partial_{\mu}{\pi^{a}}\partial^{\mu}{\pi^{a}}
- \frac{1}{2}m_{\pi}^2{\pi^{a}}^2 \nn
& -& \frac{1}{4}W_{\mu\nu}W^{\mu\nu}+\frac{1}{2}m_{\omega}^2\omega_{\mu}\omega^{\mu}
-\frac{1}{4}R^a_{\mu\nu}R^{a\mu\nu}+\frac{1}{2}m_{\rho}^2\rho^a_{\mu}\rho^{a\mu}~,
\eeq
where
\beq
&    &W_{\mu\nu}=\partial_\mu\omega_\nu-\partial_\nu\omega_\mu~,\nn
&    &R^a_{\mu\nu}=\partial_\mu\rho^a_\nu-\partial_\nu\rho^a_\mu~.
\eeq
$\psi$ is the nucleon field and $M$ the nucleon mass. Moreover, a monopole form factor is taken into account,
\beq
F_\alpha(q^2)=\frac{\Lambda^2_\alpha-m^2_\alpha}{\Lambda^2_\alpha+q^2}~,
\eeq
for each meson-nucleon vertex, where $\alpha$ is denoted to the species of meson.

Once the form factor is included into the meson exchange potential, the contact terms in the one-boson-exchange interaction become momentum dependent.  In principle, all the meson exchange interactions in Bonn potential can be expressed in the form of Yukawa functions in coordinate space even after considering the form factor effect,
\beq
V(m,r)=\frac{e^{-mr}}{r}.
\eeq
When the unitary correlation operator $U$ is applied to the two-body $NN$ interaction, we just take the following transformation in the potential $V(r)$ as,
\beq
c^\dagger(i,j) V(m,r) c(i,j) =V(m,R_+(r)).
\eeq

On the other hand, the kinetic energy part after the transformation by UCOM operator is
\beq
c^\dagger (i,j) T (i,j)-T&=&\sum_{i<j}  (\vec \alpha_i-\vec \alpha_j)\cdot \frac{\vec r}{r}\frac{1}{\sqrt{R'_+(r)}}\frac{1}{r}q_r \frac{r}{\sqrt{R'_+(r)}}\nn
&&+(\vec \alpha_i-\vec \alpha_j)\cdot \frac{\vec r}{r}\left(\frac{1}{R'_+(r)}-\frac{r}{R_+(r)}\right)q_r+\left(\frac{r}{R_+(r)}-1\right) (\vec \alpha_i-\vec \alpha_j)\cdot\vec q.
\eeq
Here, $q_r=\vec r\cdot\vec q/r$ is defined as the radial momentum.  The Dirac matrix $\vec \alpha_i$ is an operator of the Dirac spinor for $i$-th nucleon and the prime denotes differentiation with respect to the relative coordinate $r$. A trivial wave function of nuclear matter will be constructed by the Hartree-Fock wave function of free Dirac particle. Then, the expectation values of effective Hamiltonian, which is obtained by modifying the realistic $NN$ interaction with UCOM, can be evaluated. With the variational method, the Dirac equation of single nucleon is generated, which contains the scalar and vector components of self energy. After a self-consistent calculation, the energy per nucleon of nuclear matter can be calculated. Therefore, we would like to name the present framework as RHF model with UCOM (RHFU)~\cite{hu10}. 

In this work, the core of neutron star is regarded as the constituents of proton, neutron, electron, and muon, which should satisfy the charge neutrality and $\beta$ equilibrium to generate a stable structure. Their chemical potentials are constrained by the following qualities~\cite{shen02},
\beq\label{mueq}
\mu_p&=&\mu_n-\mu_e,\nn
\mu_\mu&=&\mu_e.
\eeq
The chemical potentials of nucleons and leptons are related to their Fermi surfaces at zero temperature,
\beq\label{cmq}
\mu_i&=&\sqrt{k^{i2}_F+M^{*2}_N}+U_V,\nn
\mu_l&=&\sqrt{k^{l2}_F+m^{2}_l}.
\eeq
where $U_V$ is the vector potential of nucleon in Dirac equation.
The charge neutrality requires the proton density is equal to the one of leptons,
\beq\label{nr}
\rho_p=\rho_e+\rho_\mu.
\eeq
The pressure and energy density will be obtained as a function of nucleon density with the constraints in Eqs. (\ref{mueq}) and (\ref{nr}). They are put into the TOV equation proposed by Tolman, Oppenheimer, and Volkoff to bring the properties of neutron star,
\beq\label{tov}
\frac{dP(r)}{dr}&=&-\frac{GM(r)\varepsilon(r)}{c^2r^{2}}\frac{\Big[1+\frac{P(r)}{\varepsilon(r)}\Big]\Big[1+\frac{4\pi r^{3}P(r)}{M(r)c^2}\Big]}
{1-\frac{2GM(r)}{c^2r}},\nn
\frac{dM(r)}{dr}&=&4\pi r^{2}\varepsilon(r)/c^2,
\eeq
where, $c$ is the light speed. $P(r)$ is the pressure at radius $r$ and $M(r)$ is the total mass inside a sphere of radius $r$ of neutron star.

The dimensionless tidal deformability of neutron star is defined as~\cite{fattoyev13,fattoyev18},
\beq\label{dtd}
\Lambda=\frac{2}{3}k_2C^{-5},
\eeq  
where $C=GM/Rc^2$ is the compactness parameter. $R$ and $M$ are the neutron star radius and mass, respectively. The dimensionless quadrupole tidal Love number $k_2$ is given by
\beq
k_2&=&\frac{8C^5}{5}(1-2C)^2[2+2C(y_R-1)-y_R]\nn
&&\bigg\{2C[6-3y_R+3C(5y_R-8)]\nn
&&+4C^3[13-11y_R+C(3y_R-2)+2C^2(1+y_R)]\nn
&&+3(1-2C)^2[2-y_R+2C(y_R-1)]\ln(1-2C)\bigg\}^{-1}.
\eeq
$y_R$ is a quantity which is a function value at $R$, $y(R)$. It is obtained by solving a first-order differential equation for $y$,
\beq\label{td}
r\frac{dy(r)}{dr}+y^2(r)+y(r)F(r)+r^2Q(r)=0,
\eeq 
with initial boundary condition $y(0)=2$. $F(r)$ and $Q(r)$ are the functions related to energy density, pressure, and neutron star mass,
\beq
F(r)=\bigg\{1-4\pi r^2 G[\varepsilon(r)-P(r)]/c^4 \bigg\}\left(1-\frac{2M(r)G}{rc^2}\right)^{-1},
\eeq 
and
\beq
Q(r)&=&\frac{4\pi G}{c^4}\left[5\varepsilon(r)+9P(r)+\frac{\varepsilon(r)+P(r)}{\partial P(r)/\partial\varepsilon(r)}\right]\left(1-\frac{2M(r)G}{rc^2}\right)^{-1}\nn
&&-6\left(r^2-\frac{2rM(r)G}{c^2}\right)^{-1}-\frac{4M(r)^2G^2}{r^4c^4}\nn
&&\left(1+\frac{4\pi r^3P(r)}{M(r)c^2}\right)^2\left(1-\frac{2M(r)G}{rc^2}\right)^{-2}.
\eeq
On the other hand,  the sound of speed in dense matter, $v_s$ is relevant to the derivative of pressure, $P(r)$ with respect to energy density, $\varepsilon(r)$,
\beq
\frac{\partial P}{\partial\varepsilon}=\left(\frac{v_s}{c}\right)^2.
\eeq

\section{The results and discussions}

\begin{figure}[h]
	\centering
	\includegraphics[scale=0.6]{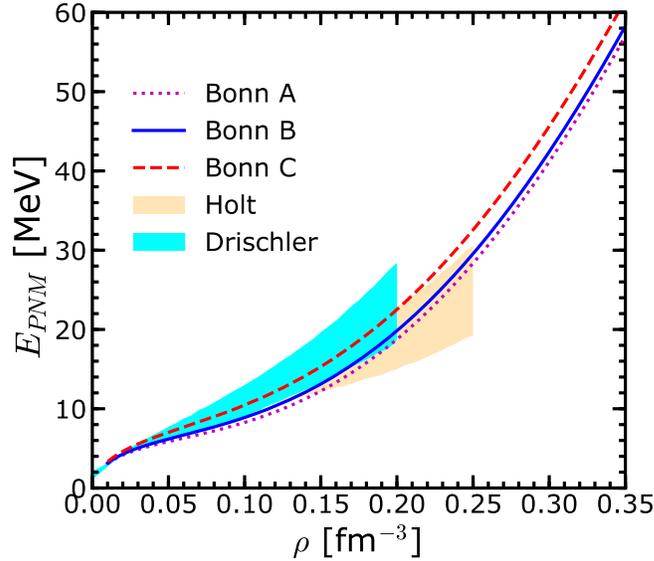}\\
	\caption{The EOSs of pure neutron matter within Bonn A, B, C potentials in the framework of RHFU model. They are compared with the results from chiral effective field theory as shown with shade regions~\cite{drischler16,holt17}. }\label{eosneu}
\end{figure}

The correlator in UCOM has the form of double exponential function, which only affect the $NN$ potentials at short-range distance. There are three parameters, $\alpha,~\beta,$ and $\eta$. They were already determined by reproducing the properties of light nuclei with few-body method and the EOS of pure neutron matter from RBHF model as $\alpha=0.8$ fm, $\beta=0.6$ fm, and $\eta=0.37$~\cite{feldmeier98,hu10}. It was found that the calculations of RHFU model also can generate the similar EOSs of neutron-rich nuclear matter by RBHF model with same UCOM correlator and same potential in our previous work. In Fig.~\ref{eosneu}, the EOSs of pure neutron matter from RHFU model with Bonn potentials are shown and compared with the latest results from the chiral effective field theory~\cite{drischler16,holt17}. They are comparable with each other.

\begin{figure}[h]
	\centering
	\includegraphics[scale=0.6]{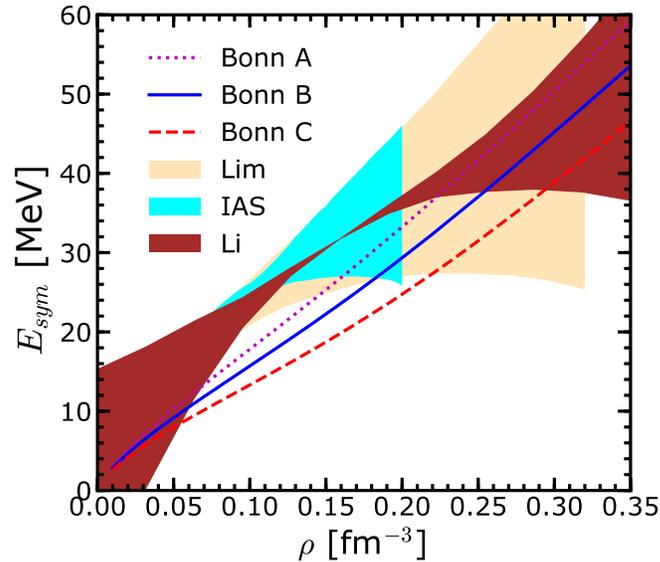}\\
	\caption{The symmetry energies as functions of density within Bonn A, B, C potentials in the framework of RHFU model and compared with several previous works
   about the constraints of symmetry energy as shown with different shade regions~\cite{baldo16,li19,lim19}.}\label{esym}
\end{figure}

 In symmetric nuclear matter, the saturation binding energy is $-14.48$ MeV at density $\rho=0.203$ fm$^{-3}$ from the Bonn A potential in the RHFU framework and the symmetry energy is $33.73$ MeV. The saturation density is a little bit far from the empirical data, which is mainly dominated by the tensor force. However in present framework, the tensor correlation is not considered~\cite{hu13}. In Fig.~\ref{esym}, the symmetry energies in RHFU model are presented as functions of density. Several previous works on the constraints of symmetry energy are also shown~\cite{baldo16,li19,lim19}. It can be found that they are comparable at low density (below $0.06$ fm$^{-3}$) and higher density (above $0.20$ fm$^{-3}$). The symmetry energy generated by the isospin symmetry plays a very important role for the neutron-rich system. The symmetry energy at empirical saturation density, $\rho_0=0.16$ fm$^{-3}$ is just $26.85$ MeV for Bonn A potential, while its most probable value is $E_\text{sym}=31.7\pm3.2$ MeV~\cite{oertel17}.  The several MeV difference is generated by the lack of tensor force. With density increasing, the short range correlation becomes more important. At $2\rho_0$, the symmetry energy in our framework is $54.00$ MeV. It is consistent with the recent ASY-EOS~\cite{russotto16} and RBHF theory~\cite{tong19}. The symmetry energies derived from Bonn B and Bonn C potentials are less than those from Bonn A potential, since their binding energies of symmetric nuclear matter are larger. Furthermore, the slope of symmetry energy at saturation density, $L$ is a very important quantity to determine the density dependence of symmetry energy. They are $96.48$ MeV, $76.86$ MeV, $47.19 $MeV, from Bonn A, B, and C potentials in our model, respectively. The symmetry energies and their slopes, $(E_\text{sym}(\rho_0), L)$ from Bonn A and Bonn B are $(33.73, 96.48)$ and $(27.00,76.86)$, respectively, which satisfy the constraint regions of symmetry energy parameters in the Fig. 2 of Ref.~\cite{lattimer13}.

\begin{figure}[h]
	\centering
	\includegraphics[scale=0.6]{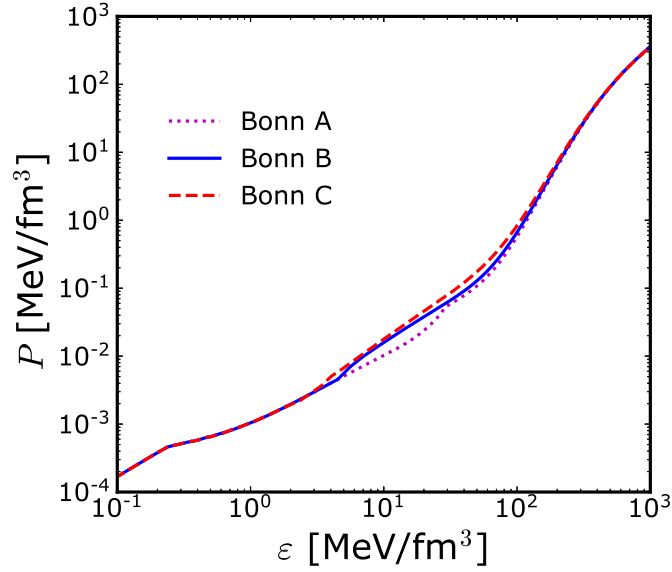}\\
	\caption{The equations of state of neutron star matter within Bonn A, B, C potentials in the framework of RHFU model.}\label{eos}
\end{figure}

Then, the EOSs of neutron star matter, i.e., $\varepsilon-P$ relation, are evaluated by using the RHFU model with Bonn potentials as shown in Fig.~\ref{eos}. In the very low density region of EOS, corresponding the crust of neutron star, there are nonuniform structures such as pasta phases. These  inhomogeneous nuclear matter can be described by the Thomas-Fermi approximation. In this work, the EOS from TM1 interaction is used, i. e. Shen EOS~\cite{shen98}, which will be matched with the EOSs from RHFU model at crust-core transition density. The EOSs from Bonn A, B, C potentials at high density are very similar. It is because that the short range correlation plays the dominant role when the distances of nucleons decrease gradually. The differences among three Bonn potentials appear from crust-core transition densities, which are about one half of saturation density, $\rho_0/2$. In this region, the role of tensor force become important~\cite{hu13}. The three Bonn potentials contained different tensor components~\cite{brockmann90}, which generate distinguished EOSs.

\begin{figure}[h]
	\centering
	\includegraphics[scale=0.6]{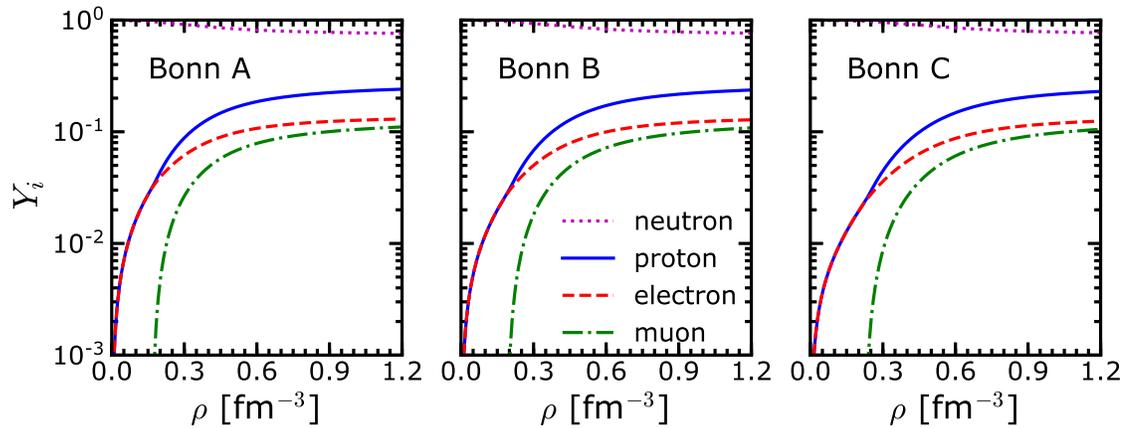}\\
	\caption{The fractions of neutron, proton, electron and muon within Bonn A, B, C potentials in the framework of RHFU model.}\label{yi}
\end{figure}

The fractions of neutron, proton, electron and muon in neutron star matter from three Bonn potentials are shown in Fig.~\ref{yi}. These fractions are very similar and become saturated at high density region. Generally speaking, the proton fractions from Bonn A potential are the largest one among three Bonn potentials due to its strongest symmetry energy which also leads to first appearance of the muon in Bonn A potential. The onset density is $0.165$ fm$^{3}$. 

\begin{figure}[h]
	\centering
	\includegraphics[scale=0.6]{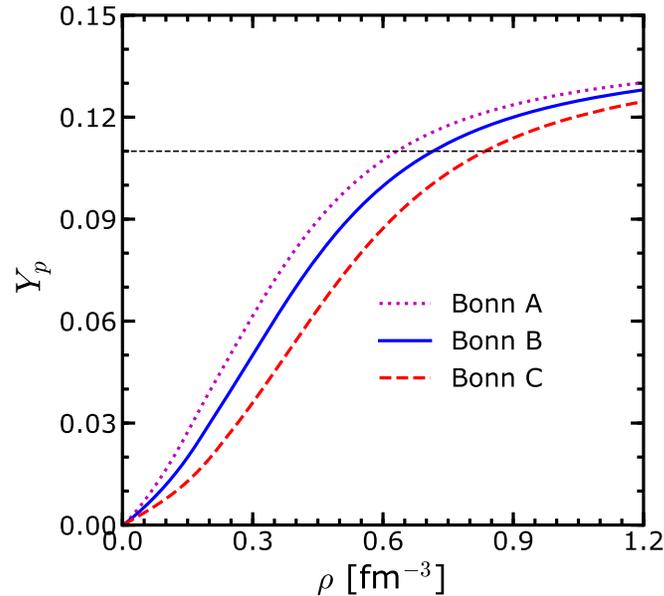}\\
	\caption{The fraction of proton within Bonn A, B, C potentials in the framework of RHFU model.}\label{yp}
\end{figure}

In the neutron star cooling, the direct Urca process plays a very important role, as $\beta$ decay of neutron and electron capture of proton. The threshold of direct Urca process is dominated by the proton fraction, $Y_p$ in neutron star. If there are only proton, neutron, and electron in neutron star, the direct Urca process will happen at $Y_p>1/9$~\cite{lattimer91}. Therefore, the proton fractions from RHFU model with three Bonn potentials are plotted in  Fig.~\ref{yp}. It can be found that the proton fraction increases slowly with density. The direct Urca process will firstly happen in Bonn A potential, where the threshold density is about $0.6$ fm$^{-3}$. This threshold density is postponed in Bonn B and Bonn C potentials. Actually, the proton fraction, $Y_p$ is related with the symmetry energy, which can be approximated as the cube of symmetry energy in neutron star~\cite{lattimer16}.

\begin{figure}[h]
	\centering
	\includegraphics[scale=0.6]{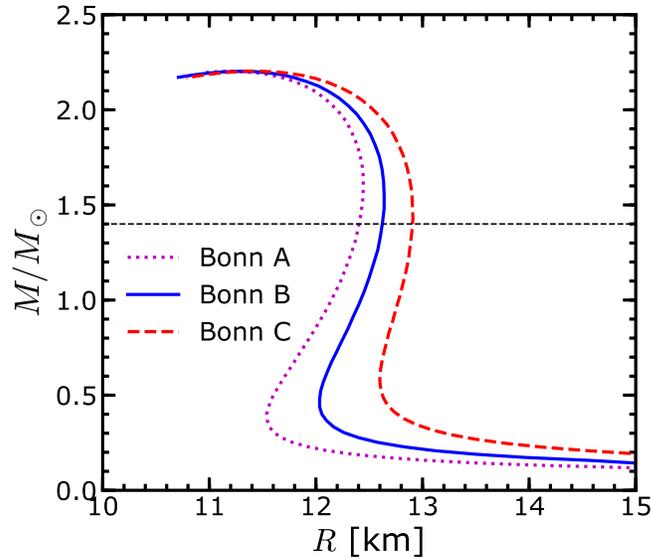}\\
	\caption{The mass-radii relations of neutron star withing Bonn A, B, C potentials.}\label{rmn}
\end{figure}

\begin{figure}[h]
	\centering
	\includegraphics[scale=0.6]{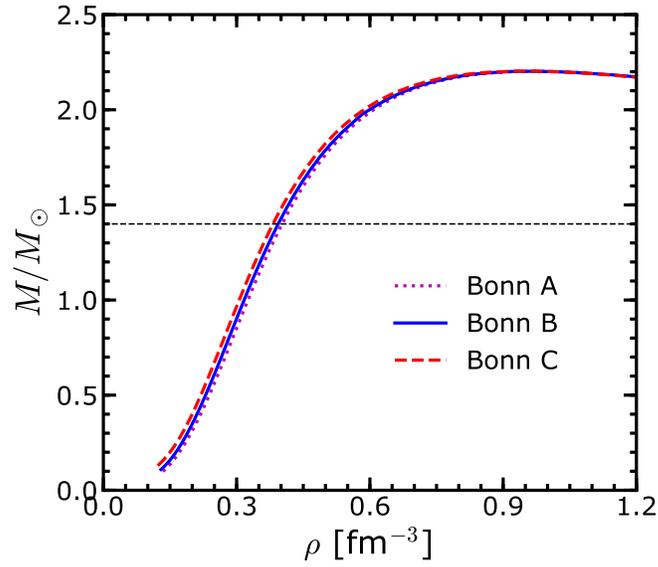}\\
	\caption{The mass-density relations of neutron star withing Bonn A, B, C potentials.}\label{rhom}
\end{figure}

Once the EOSs are obtained, the global properties of neutron star can be calculated by solving TOV equation. The mass-radii relation and mass-density relations are shown in Figs.~\ref{rmn} and \ref{rhom}. The maximum masses of neutron star from Bonn potentials are all around $2.2 M_\odot$ and the corresponding radius are $11.18-11.48$ km, which are consistent with results from RBHF model and relativistic central variational method with same potentials~\cite{krastev06,hu17}. It is naturally to understand that the maximum masses of neutron star are same, since three Bonn potentials generate the analogous EOSs at high density. However, the differences of EOS in the intermediate density will influence the mass-radii relation of lower neutron star mass, that are measured in the observations of pulsar star at most. The radius at $1.4 M_\odot$, $R_{1.4}$ , from Bonn A, B, C potentials are $12.40, ~12.62$ and $12.90$ km, respectively. These radius are consistent with the latest constraints from GW170817, where $R_{1.4}$ should be less than $13.8$ km~\cite{fattoyev18,lim18,baiotti19}. The mass-density relations among three Bonn potentials are very similar. The slight differences arise at low density region due to the EOSs. The central densities of the maximum mass are around $0.95$ fm$^{-3}$, that are higher than those in RMF model, like TM1 parameter set. These densities reduce to  $0.40$ fm$^{-3}$ for $1.4 M_\odot$ neutron star.

\begin{figure}[h]
	\centering
	\includegraphics[scale=0.6]{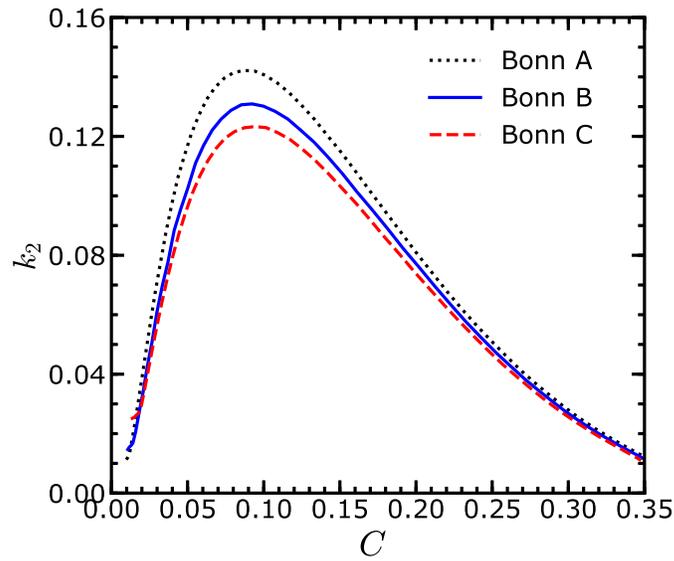}\\
	\caption{The second Love number as a function of compactness from Bonn A, B, C potentials.}\label{rmk}
\end{figure}

When there is a collision between two neutron stars, the tidal deformability of a star in an external gravitational field from another star can be extracted from the gravitational wave of two neutron star merger. Therefore, the tidal deformability of neutron star is growing into another new constraint to the EOS, besides its mass and radii. As shown in Eq.(\ref{dtd}), the dimensionless tidal deformability is related to the compactness parameter and the dimensionless second Love number, which is obtained by solving a differential equation related to the pressure, energy density, mass, and the speed of sound. The second Love number as a function of compactness from Bonn A, B, C potentials are plotted in Fig.~\ref{rmk}. The second Love number firstly increases with compactness parameter rapidly, reaches the maximum around $C=0.09$ and then decreases slowly. The maximum values of second Love number from Bonn A, B, C potentials are $0.142,~0.131$, and $0.123$, respectively, which are consistent on the other methods, like RMF an SHF frameworks~\cite{malik18}.

\begin{figure}[h]
	\centering
	\includegraphics[scale=0.6]{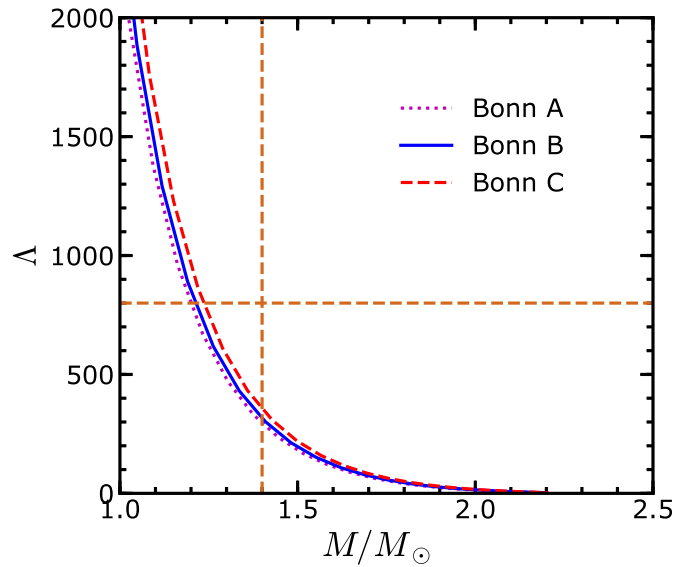}\\
	\caption{The tidal deformability as a function of neutron star mass from Bonn A, B, C potentials.}\label{rml}
\end{figure}

After the second Love number, $k_2$ calculated, the dimensionless tidal deformability, $\Lambda$ can be immediately obtained, which are shown in Fig.~\ref{rml} as a function of neutron star mass within the RHFU model. It rapidly reduces with neutron star mass increasing and approaches to zero at the maximum neutron star mass. When compactness parameter, $C$ is small, the value of $k_2$ is also very small. Due to $\Lambda\propto C^{-5}$, therefore at small neutron mass, the $\Lambda$ is very large. Inversely, at large neutron mass region, the $C$ is a finite number, while $k_2$ becomes very small again, which generates a very small $\Lambda$. Furthermore, the $\Lambda$ from Bonn C potential is largest. It is because that the radii of neutron star from Bonn C is the largest one at a fixing neutron star mass. In this case, $\Lambda\propto R^{5}$. The masses of neutron star in GW170817 event were regarded to around $1.4M_\odot$. Therefore, the dimensionless tidal deformability at $1.4M_\odot$, $\Lambda_{1.4}$, is a more useful quantity to constrain the EOS. There were already many works to analyze the data from GW170817 to extract the value of $\Lambda_{1.4}$. 
For example, it was estimated as $\Lambda_{1.4}=190^{+390}_{-120}$ in the latest analysis from LIGO and Virgo Collaborations~\cite{abbott18a}. De {\it et al.} also gave the binary tidal deformability, $\tilde \Lambda=245^{+453}_{-151}$ of GW170817, with a Bayesian estimation, where the component mass prior was generated by the radio observations of double neutron stars~\cite{de18}.
In this work, the $\Lambda_{1.4}$  from Bonn A, B, C potentials are $292,~318$, and $354$, respectively. These values are completely consistent with the present constraint about $\Lambda_{1.4}$.

\begin{figure}[h]
	\centering
	\includegraphics[scale=0.6]{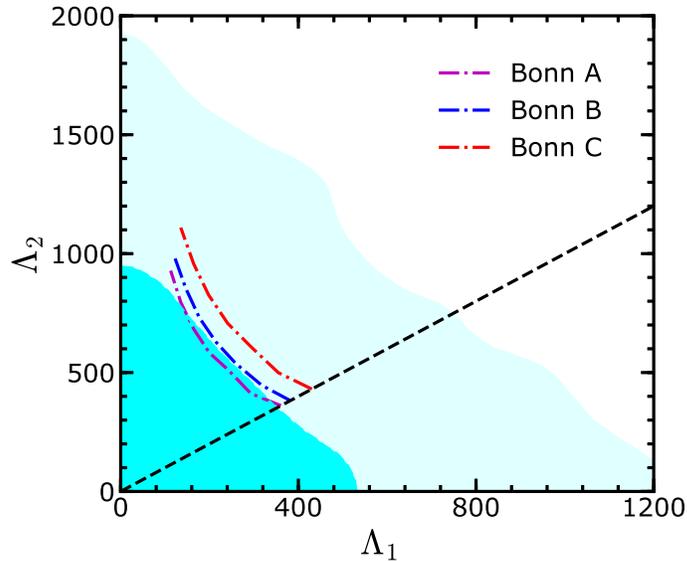}\\
	\caption{The tidal deformabilities of the binaries in GW170817 event from Bonn A, B, C potentials, where the light and deep shadow regions ar the 90\% and 50\% credible intervals from the GW170817, respectively.}\label{rmll}
\end{figure}

In GW170817 event, the chirp mass $\mathcal{M}=(m_1m_2)^{3/5}(m_1+m_2)^{-1/5}$ was measured, where $m_1$ and $m_2$ are the masses of two merger neutron stars. However, the accurate values of these two neutron stars cannot be obtained from the gravitation-wave signal. There is only possible distribution. Therefore, it is assumed that the heavier neutron star is in the range from $1.365M_\odot$ to $1.600M_\odot$. The mass of lighter one is from $1.170M_\odot$ to $1.365M_\odot$ with the constraint of chirp mass. The relations between their tidal deformabilities, $\Lambda_1(m_1)$ and  $\Lambda_2(m_2)$ is given in Fig.~\ref{rmll} from Bonn A, B, C potentials. The $90\%$ and $50\%$ credible region for the tidal deformabilities of two neutron stars from the analysis of GW170817 for the low-spin prior are presented as light cyan area and dark cyan area, respectively~\cite{abbott18a}. It can be found that the tidal deformabilities of the binary stars from RHFU model are located in the constraint region from GW170817.

\section{Conclusions}
The properties of neutron star were investigated by the relativistic Hartree-Fock model with unitary correlation operator method (RHFU). The realistic $NN $ interaction, Bonn potentials were used. Their strong repulsions at short range distance were taken into account by unitary correlation operator method (UCOM), while the tensor correlation was neglected due the neutron-rich environment at neutron star. Therefore, the neutron star matter can be reasonably described by the RHFU model.

The equations of state (EOSs) of neutron star matter were obtained in the conditions of $\beta$ equilibrium and charge neutrality with Bonn A, B, C potentials. Their differences are mainly located at intermediate density region, where the tensor force plays a very important role, while the tensor components of three Bonn potentials are distinguished. With density increasing, the EOSs are almost identical. At the dense matter, the short range correlation of $NN$ potential was more essential. The proton fraction is related with the direct Urca process in the neutron star cooling. The threshold of direct Urca process happened earliest in Bonn A potential, which had the largest symmetry energy.   

The maximum masses of neutron star from all three Bonn potentials were around $2.2M_\odot$. The corresponding radius were $11.18-11.48$ km, which are consistent with the calculations in RBHF mode and relativistic central variational method by using the same potentials, while the radius at $1.4M_\odot$ were $12.40-12.90$ km. These values satisfied the constraint from GW170817. The dimensionless tidal deformability, $\Lambda$ of neutron star was also studied as a function of neutron star mass in RHFU model. They were $292,~318$ and $354$ at $1.4M_\odot$ for Bonn A, B, C potentials, respectively. A lot of analysis works for GW170817 event indicated that $\Lambda_{1.4}$ should be less than $800$. The tidal deformabilities of binaries in GW170817 from RHFU model were also located in the $90\%$ credible interval extracted directly by LIGO and Virgo Collaborations.

The calculations of RHFU model for neutron star are consistent with the results from RBHF model, which demonstrate that the RHFU model is a very powerful tool to study the neutron-rich system in a very simple framework. It is necessary to include the tensor correlation in RHFU model, so that it can be applied to more aspects in nuclear physics within the realistic $NN$ potential. 

\section{Acknowledgments}
This work was supported in part by the National Natural Science Foundation of China (Grants No. 11405116 and No. 11775119).


\begin{thebibliography}{99}
	\bibitem{abbott17a} B. P. Abbott {\it et al.}, (LIGO Scientific and Virgo Collaboration), Phys. Rev. Lett. {\bf 119}, 161101 (2017).
	\bibitem{abbott17b} B. P. Abbott {\it et al.}, (Virgo, Fermi-GBM, INTEGRAL, and LIGO Scientific Collaboration), Astrophys. J. {\bf 848}, L13 (2017).
	\bibitem{abbott17c} B. P. Abbott {\it et al.}, Astrophys. J. {\bf 848}, L12 (2017).
	\bibitem{goldstein17} A. Goldstein {\it et al.}, Astrophys. J. {\bf 848}, L14 (2017).
	\bibitem{coulter17} D. A. Coulter {\it et al.}, Science {\bf358}, 1556 (2017).
	\bibitem{troja17} E. Troja {\it et al.}, Nature (London) {\bf 551}, 71 (2017).
	\bibitem{haggard17} D. Haggard, M. Nynka, J. J. Ruan, V. Kalogera, S. Bradley Cenko, P. Evans, and J. A. Kennea, Astrophys. J. {\bf 848}, L25 (2017).
	\bibitem{abbott18a} B. P. Abbott {\it et al.}, (LIGO Scientific and Virgo Collaboration), Phys. Rev. Lett. {\bf 121}, 161101 (2018).
	\bibitem{hinderer08}T. Hinderer, Astrophys. J. {\bf 677}, 1216 (2008); {\bf 697}, 964(E)(2009).
	\bibitem{hinderer10}T. Hinderer, B. D. Lackey, R. N. Lang, and J. S. Read, Phys. Rev. D {\bf 81}, 123016 (2010).
	\bibitem{fattoyev13}F. J. Fattoyev, J. Carvajal, W. G. Newton, and B.-A. Li, Phys. Rev. C {\bf 87}, 015806 (2013).
	\bibitem{fattoyev18}F. J. Fattoyev, J. Piekarewicz, and C. J. Horowitz, Phys. Rev. Lett. {\bf 120}, 172702 (2018).
	\bibitem{annala18} E. Annala, T. Gorda, A. Kurkela, and Aleksi Vuorinen, Phys. Rev. Lett. {\bf 120}, 172703 (2018).
	\bibitem{most18} E. R. Most, L.R. Weih, L. Rezzolla, and J. Schaffner-Bielich, Phys. Rev. Lett. {\bf 120}, 261103 (2018).
	\bibitem{lim18} Y. Lim and J. W. Holt, Phys. Rev. Lett. {\bf 121}, 062701 (2018).
	\bibitem{tews18} I. Tews, J. Margueron, and S. Reddy, Phys. Rev. C {\bf 98}, 045804 (2018).
	\bibitem{zhang18} N. B. Zhang, B. A. Li, and J. Xu, Astrophys. J. {\bf 859}, 90 (2018).
	\bibitem{oertel17} M. Oertel, M. Hempel, T. Kl\"{a}hn, and S. Typel. Rev. Mod. Phys. {\bf 89}, 015007 (2017).
	\bibitem{shen02} H. Shen, Phys. Rev. C {\bf 65}, 035802 (2002).
	\bibitem{vautherin72} D. Vautherin and D. M. Brink, Phys. Rev. C {\bf 5}, 626 (1972).
	\bibitem{bender03} M. Bender, P. H. Heenen, and P. G. Reinhard, Rev. Mod. Phys. {\bf 75}, 121 (2003).
	\bibitem{stone07} J. R. Stone and P. G. Reinhard, Prog. Part. Nucl. Phys. {\bf 58}, 587 (2007).
	\bibitem{goriely09} S. Goriely, S. Hilaire, M. Girod, and S. P\'{e}ru, Phys. Rev. Lett. {\bf 102}, 242501 (2009).
	\bibitem{serot86} B. D. Serot and J. D. Walecka,  Adv. Nuc. Phys. {\bf 16}, 1 (1986).
	\bibitem{ring96} P. Ring, Prog. Part. Nucl. Phys. {\bf 37}, 193 (1996).
	\bibitem{meng06} J. Meng, H. Toki, S. G. Zhou, S. Q. Zhang, W. H. Long, and L. S. Geng, Prog. Part. Nucl. Phys. {\bf 57}, 470 (2006). 
	\bibitem{liang15} H. Z. Liang, J. Meng, and S. G. Zhou, Phys. Rep. {\bf 570}, 1 (2015).
	\bibitem{long06}W. H. Long, N. Van Giai, and J. Meng, Phys. Lett. B {\bf 640}, 150 (2006).
	\bibitem{long07}W. H. Long, H. Sagawa, N. Van Giai, and J. Meng, Phys. Rev. C {\bf 76}, 034314 (2007).
	\bibitem{sun08} B. Y. Sun, W. H. Long, J. Meng, and U. Lombardo, Phys. Rev. C {\bf 78}, 065805 (2008).
	\bibitem{long12} W. H. Long, B. Y. Sun, K. Hagino, and H. Sagawa, Phys. Rev. C {\bf 85}, 025806 (2012).
	\bibitem{erler12} J. Erler {\it et al.}, Nature {\it 486}, 509 (2012)
	\bibitem{xia18} X. W. Xia, Y. Lim, P. W. Zhao, H. Z. Liang, X. Y. Qu, Y. Chen, H. Liu, L. F. Zhang, S. Q. Zhang, Y. Kim, and J. Meng, Atom. Data Nucl. Data Tables, {\bf 121}, 1 (2018).
	\bibitem{dutra12} M. Dutra, O. Lourenco, J. S. S\'a Martins, A. Delfino, J. R. Stone, and P. D. Stevenson, Phys. Rev. C {\bf 85}, 035201 (2012).
	\bibitem{dutra14} M. Dutra, O. Lourenco, S. S. Avancini, B. V. Carlson, A. Delfino, D. P. Menezes, C. Provid\^{e}ncia, S. Typel, and J. R. Stone, Phys. Rev. C {\bf 90}, 055203 (2014).
	\bibitem{machleidt87} R. Machleidt, K. Holinde, and Ch. Elster, Phys. Rep. {\bf 149}, 1 (1987).
	\bibitem{machleidt89} R. Machleidt, Adv. Nucl. Phys.  {\bf 19}, 189 (1989). 
	\bibitem{wiringa95} R. B. Wiringa, V. G. J. Stoks, and R. Schiavilla, Phys. Rev. C {\bf 51}, 38 (1995).
	\bibitem{machleidt01} R. Machleidt, Phys. Rev. C {\bf 63}, 024001 (2001).
	\bibitem{epelbaum15a} E. Epelbaum, H. Krebs, and U.-G. Mei\ss ner, Eur. Phys. J. A {\bf 51}, 53 (2015).
	\bibitem{epelbaum15b} E. Epelbaum, H. Krebs, and U.-G. Mei\ss ner, Phys. Rev. Lett. {\bf 115}, 122301 (2015).
	\bibitem{entem15} D. R. Entem, N. Kaiser, R. Machleidt, and Y. Nosyk, Phys. Rev. C {\bf 91}, 014002 (2015).
	\bibitem{akmal98} A. Akmal, V. R. Pandharipande, and D. G. Ravenhall,  Phys. Rev. C {\bf 58}, 1804 (1998).
	\bibitem{carlson15}J. Carlson  {\it et al.}, Rev. Mod. Phys. {\bf 87}, 1067 (2015).
	\bibitem{dickhoff04}W. H. Dickhoff and C. Barbieri, Prog. Part. Nucl. Phys. {\bf 52}, 377 (2004).
	\bibitem{hagen14a} G. Hagen, T. Papenbrock, M. Hjorth-Jensen, and D. J. Dean, Rep. Prog. Phys. {\bf 77}, 096302 (2014).
	\bibitem{carbone13} A. Carbone, A. Polls, and A. Rios, Phys. Rev. C {\bf 88}, 044302 (2013).
	\bibitem{carbone14} A. Carbone, A. Rios, and A. Polls, Phys. Rev. C {\bf 90}, 054322 (2014).
	\bibitem{hagen14} G. Hagen, T. Papenbrock, A. Ekstr\"{o}m, K. A. Wendt, G. Baardsen, S. Gandolfi, M. Hjorth-Jensen, and C. J. Horowitz, Phys. Rev. C {\bf 89}, 014319 (2014).
	\bibitem{drischler14} C. Drischler, V. Som$\grave{a}$, and A. Schwenk, Phys. Rev. C {\bf 89}, 025806 (2014).
	\bibitem{drews15} M. Drews and W. Weise, Phys. Rev. C {\bf 91}, 035802 (2015).
	\bibitem{drews16} M. Drews and W. Weise, Prog. Part. Nucl. Phys. in press, (2016).
	\bibitem{li06} Z. H. Li, U. Lombardo, H.-J. Schulze, W. Zuo, L. W. Chen, and H. R. Ma,  Phys. Rev. C {\bf 74}, 047304 (2006).
	\bibitem{baldo07} M. Baldo, C. Maieron, J. Phys. G {\bf 34}, R243 (2007).
	\bibitem{brockmann90} R. Brockmann and R. Machleidt,  Phys. Rev. C {\bf 42}, 1965 (1990).
	\bibitem{krastev06} P. G. Krastev and F. Sammarruca,  Phys. Rev. C {\bf 74}, 025808 (2006).
	\bibitem{sammarruca10} F. Sammarruca, Int. J. Mod. Phys. E {\bf 19}, 1259 (2010).
	\bibitem{dalen10} E. N. E. Dalen and H. Muether, Int. J. Mod. Phys. E {\bf 19}, 2077 (2010).
	\bibitem{shen19} S. H. Shen, H. Z. Liang, W. H. Long, J. Meng, and P. Ring, Prog. Part. Nucl. Phys., in press
	\bibitem{demorest10} P. B. Demorest, T. Pennucci, S. M. Ransom, M. S. E. Roberts, and J. W. T. Hessels, Nature (London) \textbf{467}, 1081 (2010).
	\bibitem{fonseca16} E. Fonseca {\it et al.}, Astrophys. J. {\bf 832}, 167 (2016).
	\bibitem{antoniadis13} J. Antoniadis, P. C. C. Freire, N. Wex, T. M. Tauris, R. S. Lynch \textit{et al.}, Science \textbf{340}, 6131 (2013).
	\bibitem{cromartie19} H. T. Cromartie {\it et al.}, arXiv:1904.06759.
	\bibitem{baiotti19} L. Baiotti, arXiv:1907.08534.
	\bibitem{tong19}H. Tong, P. W. Zhao, and J. Meng, arXiv:1903.05938.
	\bibitem{wang12} Y. Wang, J. Hu, H. Toki, and H. Shen, Prog. Theo. Phys. {\bf 127}, 739 (2012).
	\bibitem{hu13} J. Hu, H. Toki, and Y. Ogawa, Prog. Theor. Exp. Phys. {\bf 103D02}, (2013).
	\bibitem{hu11} J. Hu, H. Toki, and H. Shen, J. Phys. G. {\bf 38}, 085105 (2011).
	\bibitem{hu17} J. Hu, H. Shen, and H. Toki,  Phys. Rev. C {\bf 95}, 025804 (2017).
	\bibitem{hu10} J. Hu, H. Toki, W. Wen, and H. Shen,  Phys. Lett. B {\bf 687}, 271 (2010).
	\bibitem{feldmeier98} H. Feldmeier, T. Neff, and R. Roth, J. Schnack, Nucl. Phys. A {\bf 632}, 61 (1998).
	\bibitem{neff03} T. Neff and H. Feldmeier, Nucl. Phys. A {\bf 713}, 311 (2003).
	\bibitem{drischler16} C. Drischler, A. Carbone, K. Hebeler, and A. Schwenk, Phys. Rev. C {\bf 94}, 054307 (2016).
	\bibitem{holt17} J. W. Holt and N. Kaiser, Phys. Rev. C {\bf 95}, 034326 (2017).
    \bibitem{baldo16} M. Baldo and G. F. Burgio, Prog. Part. Nucl. Phys. {\bf 91}, 203 (2016).
    \bibitem{li19} B. A. Li, P. G. Krastev, D. H. Wen, and N. B. Zhang,  Eur. Phys. J. A {\bf 55},117 (2019).
    \bibitem{lim19} Y. Lim and J. W. Holt, arXiv:1902.05502.
	\bibitem{russotto16} P. Russotto {\it et al.}, Phys. Rev. C {\bf 94}, 034608 (2016).
	 \bibitem{lattimer13} J. M. Lattimer and Y. Lim,  Astrophys. J. {\bf 771}, 51 (2013).
	\bibitem{shen98} H.Shen, H.Toki, K.Oyamatsu, and K.Sumiyoshi, Nucl. Phys. A {\bf 637}, 435 (1998).
	\bibitem{lattimer91} J. M. Lattimer, C. J. Pethick, M. Prakash, and P. Haensel, Phys. Rev. Lett. {\bf 66}, 2701 (1991).
	\bibitem{lattimer16}J. M. Lattimer and M. Prakash, Phys. Rep. {\bf 621}, 127 (2016).
	\bibitem{malik18}T. Malik, N. Alam, M. Fortin, C. Provid\^{e}ncia, B. K. Agrawal, T. K. Jha, B. Kumar, and S. K. Patra, Phys. Rev. C {\bf 98}, 035804 (2018).
	\bibitem{de18} S. De, D. Finstad, J. M. Lattimer, D. A. Brown, E. Berger, and C. M. Biwer, Phys. Rev. Lett. {\bf 121}, 091102 (2018).
	
	
	

\end{thebibliography}
\end{document}